\documentclass[prl,twocolumn,letterpaper]{revtex4-1}
\usepackage{bm,graphicx,graphics,amsmath,amssymb,bm,epsfig,color}
\usepackage{euscript,tabularx}
\usepackage{longtable}

\begin{document}

\title{Origin of spectral purity and tuning sensitivity \\ in a
  vortex-based spin transfer nano-oscillator}

\author{A. Hamadeh}
\author{G. de Loubens}
\email{gregoire.deloubens@cea.fr}
\author{O. Klein}

\affiliation{Service de Physique de l'\'Etat Condens\'e (CNRS URA
  2464), CEA Saclay, 91191 Gif-sur-Yvette, France}

\author{V.V. Naletov}

\affiliation{Service de Physique de l'\'Etat Condens\'e (CNRS URA
  2464), CEA Saclay, 91191 Gif-sur-Yvette, France}
\affiliation{Unit\'e Mixte de Physique CNRS/Thales and Universit\'e
  Paris Sud 11, 1 av. Fresnel, 91767 Palaiseau, France}
\affiliation{Institute of Physics, Kazan Federal University, Kazan
    420008, Russian Federation}

\author{N. Locatelli}
\author{R. Lebrun}
\author{J. Grollier}
\author{V. Cros}

\affiliation{Unit\'e Mixte de Physique CNRS/Thales and Universit\'e
  Paris Sud 11, 1 av. Fresnel, 91767 Palaiseau, France}

\date{\today}

\begin{abstract}

  We investigate the microwave characteristics of a spin transfer
  nano-oscillator (STNO) based on coupled vortices as a function of
  the perpendicular magnetic field $H_\perp$. While the generation
  linewidth displays strong variations on $H_\perp$ (from 40~kHz to
  1~MHz), the frequency tunability in current remains almost constant
  ($\simeq 7$~MHz/mA). We demonstrate that our vortex-based oscillator
  is quasi-isochronous independently of $H_\perp$, so that the severe
  nonlinear broadening usually observed in STNOs does not
  exist. Interestingly, this does not imply a loss of frequency
  tunability, which is here governed by the current induced Oersted
  field. Nevertheless this is not sufficient to achieve the highest
  spectral purity in the full range of $H_\perp$ either: we show that
  the observed linewidth broadenings are due to the excited mode
  interacting with a lower energy overdamped mode, which occurs at the
  successive crossings between harmonics of these two modes.  These
  findings open new possibilities for the design of STNOs and the
  optimization of their performance.

\end{abstract}

\maketitle

A spin-polarized current exerts on a ferromagnetic material a torque
that can compensate the damping and lead to auto-oscillation of the
magnetization \cite{slonczewski96,berger96,kiselev03,rippard04}. Owing
to their tunability, agility, compactness and integrability, spin
transfer nano-oscillators (STNOs) are promising candidates for various
high frequency applications such as frequency generation, signal
processing and microwave frequency detection. Spectral purity and
tuning sensitivity are two key characteristics for such devices. A
particularity of STNOs compared to other oscillators is their strong
nonlinear properties, which are inherited from the equation of motion
of magnetization \cite{slavin09}. On one hand, they confer interesting
properties to STNOs, as for instance their large frequency tunability
\cite{rippard04,hamadeh12}. On the other hand, they lead to a severe
broadening of the generation linewidth \cite{kim08}, which is the main
limiting factor to their practical applications. In this Letter, we
shall demonstrate that strong nonlinearities are not necessary to
achieve large tuning sensitivity, while weak nonlinearities are not
sufficient to obtain high spectral purity, which points towards
alternative routes to engineer STNOs with improved performance.

\begin{figure}
  \includegraphics[width=\columnwidth]{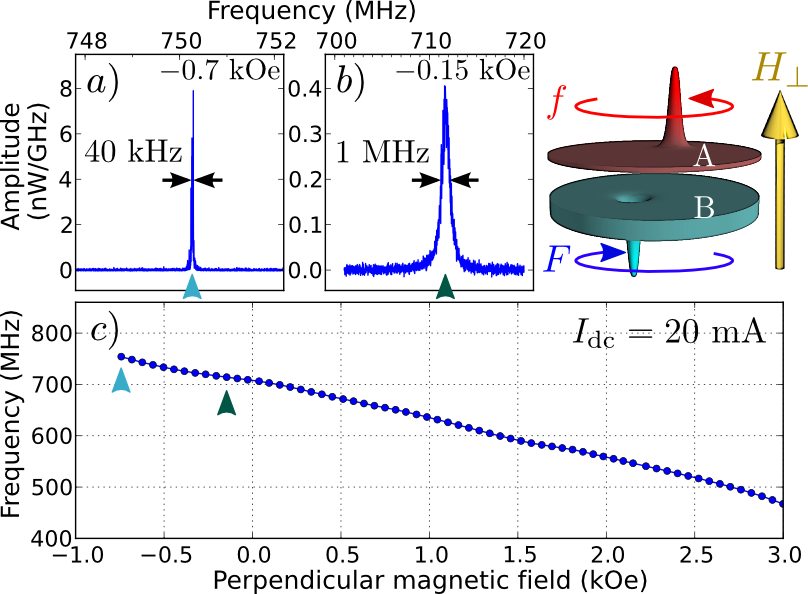}
  \caption{(Color online) Upper right sketch: STNO based on coupled
    vortices. (a) Power density spectra measured at
    $I_\text{dc}=20$~mA and a perpendicular magnetic field
    $H_\perp=-0.7$~kOe and (b) $H_\perp=-0.15$~kOe. (c) Dependence of
    the fundamental oscillation frequency on $H_\perp$.}
  \label{fig:1}
\end{figure}

\begin{figure}
  \includegraphics[width=\columnwidth]{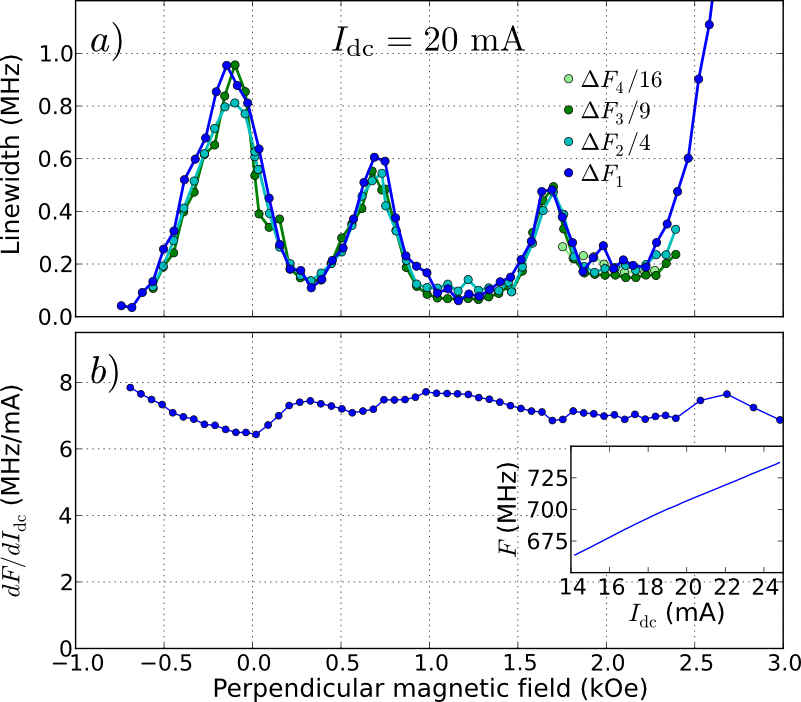}
  \caption{(Color online) (a) Generation linewidths of harmonics $n=1$
    to 4 divided by $n^2$ as a function of $H_\perp$. (b) Frequency
    tunability vs. $H_\perp$. Inset: dependence of the oscillation
    frequency on $I_\text{dc}$ measured at $H_\perp=0$~kOe.}
  \label{fig:2}
\end{figure}

So far, some of the best microwave characteristics have been reported
for STNOs in which spin transfer torque (STT) excites the gyrotropic
mode of a magnetic vortex \cite{pribiag07}. It results in microwave
emission in the range 100~MHz$-$2~GHz characterized by a narrow
linewidth (about 1~MHz) and large output power in the case of TMR
devices \cite{dussaux10}. Moreover, the oscillation frequency of
vortex-based STNOs can be rapidly switched between different values
using the bias current, which demonstrates their high agility and
tunability in current \cite{manfrini09}. Recently, we have reported a
record high spectral purity (quality factor $Q>15000$) in a spin-valve
nano-pillar where STT drives the dynamics of two coupled vortices (one
in each ferromagnetic layer) \cite{locatelli11}. In such a STNO, the
generation linewidth at fixed dc current $I_\text{dc}$ displays strong
variations (down from 40~kHz [Fig.\ref{fig:1}a] up to 1~MHz
[Fig.\ref{fig:1}b]) on the applied perpendicular field $H_\perp$, as
displayed in Fig.\ref{fig:2}a. At the same time, its frequency
tunability $dF/dI_\text{dc} \simeq 7$~MHz/mA remains almost constant
(see Fig.\ref{fig:2}b), which points out that spectral purity and
tuning sensitivity are uncorrelated. In the following, we aim at
understanding the physical origin of these peculiar behaviors, as this
will give hints to optimize the characteristics of STNOs.

The studied STNO is a circular nanopillar of diameter $2R=250$~nm
patterned from a (Cu60$|$Py$_\text{B}$15
$|$Cu10$|$Py$_\text{A}$4$|$Au25) stack, where thicknesses are in nm
and Py=Ni$_{80}$Fe$_{20}$, which can sustain the double vortex
configuration \cite{locatelli11,sluka12}. A current $I_\text{dc}>0$ is
injected through the STNO using the bottom Cu and top Au electrodes,
which corresponds to electrons flowing from the thick Py$_\text{B}$ to
the thin Py$_\text{A}$ layer. In both Py layers, the vortex
chiralities are parallel to the orthoradial Oersted field produced by
$I_\text{dc}$. We use a magnetic field $H_\perp$ perpendicular to the
sample plane in order to control the relative orientation of the
vortex core polarities. As shown in Ref.\cite{locatelli11}, we observe
a narrow microwave signal only in the case when they are opposite,
$p_\text{A}p_\text{B}=-1$, and when $I_\text{dc}$ exceeds a threshold
current $I_\text{th}\simeq8$ to 11~mA depending on $H_\perp$. The
spontaneous microwave emission branch at $I_\text{dc}=20$~mA for which
$p_\text{A}=+1$ and $p_\text{B}=-1$ is displayed in
Fig.\ref{fig:1}c. The oscillation frequency linearly decreases with
increasing $H_\perp$, as expected for a gyrotropic mode dominated by
the thick layer vortex, whose polarity is antiparallel to the applied
field \cite{loubens09}. The observed emission frequency, which
decreases from 750~MHz to 450~MHz as $H_\perp$ increases from
$-0.7$~kOe to 3~kOe, agrees well \cite{locatelli11} with the expected
gyrotropic frequency for the thick Py$_\text{B}$ layer augmented with
the contributions of the Oersted field \cite{khvalkovskiy09,dussaux12}
and of the dipolar coupling to the thin Py$_\text{A}$ layer
\cite{guslienko05a}. The boundaries of the frequency branch shown in
Fig.\ref{fig:1}c result from the combined action of $H_\perp$ and STT
to reverse the vortex cores in the Py layers \cite{locatelli13}.

We now concentrate on the dependence of the generation linewidth on
$H_\perp$, which is plotted in Fig.\ref{fig:2}a using dark blue
symbols ($\Delta F_1$). It displays minima down to 40~kHz ($F_1/\Delta
F_1 \simeq 19000$) and maxima up to 1~MHz ($F_1/\Delta F_1 \simeq
700$), \textit{i.e.}, a decrease of spectral purity by more than a
factor 25 by changing $H_\perp$ from $-0.7$~kOe [Fig.\ref{fig:1}a] to
$-0.15$~kOe [Fig.\ref{fig:1}b]. We first note that the strong
variations of the linewidth observed at $I_\text{dc}=20$~mA cannot be
attributed to changes of the nearly constant supercriticality
($I_\text{dc}/I_\text{th} \simeq 2$). The generation linewidth of a
nonlinear auto-oscillator can be written as \cite{kim08}:
\begin{equation}
  \Delta F_\text{g} = \frac{k_{B}T}{E_s} \frac{\Gamma_+}{2\pi} (1+\nu^2) \, ,
  \label{eq:deltaf}
\end{equation}
where $k_\text{B}$ is the Boltzmann constant, $T$ the temperature,
$E_s$ the energy stored in the auto-oscillation, $\Gamma_+$ the
natural energy dissipation rate, and $\nu$ the nonlinear
phase-amplitude coupling parameter. The latter is defined as $\nu =
Np/\Gamma_p$, the dimensionless ratio between the nonlinear frequency
shift $N$ multiplied by the normalized oscillation power $p$ and
divided by the damping rate of amplitude fluctuations $\Gamma_p$
\cite{slavin09}.

Since the spectral linewidth $\Delta F_\text{g}$ depends quadratically
on $\nu$, the dependence of this parameter on $H_\perp$ should be
evaluated. For this, we analyze the linewidth of the harmonics of the
auto-oscillation signal. It was shown in Ref.\cite{quinsat12} that the
linewith $\Delta F_n$ of the $n^\text{th}$ harmonics is related to the
fundamental one ($\Delta F_1$) by:
\begin{equation}
  \frac{1}{n^2}=\Delta F_1\left(\frac{1+\nu^2}{\Delta F_n}-\nu^2 \frac{1-\exp{(-2\Gamma_p/\Delta F_n)}}{2\Gamma_p}\right) \, .
  \label{eq:harm}
\end{equation}
We have plotted in Fig.\ref{fig:2}a the evolution of $\Delta F_n/n^2$
($n=2$ to 4) together with the one of $\Delta F_1$. It is clear from
this graph that independently of $H_\perp$, $\Delta F_n \simeq n^2
\Delta F_1$, which means that our STNO is quasi-isochronous and from
Eq.(\ref{eq:harm}), that $\nu \simeq 0$ in the full field
range. Therefore, one can also exclude that the strong variations of
linewidth observed in Fig.\ref{fig:2}a are due to some changes of
$\nu$ with $H_\perp$. Moreover, we can estimate the generation
linewidth from Eq.(\ref{eq:deltaf}) when $\nu=0$. For the vortex
gyrotropic mode with angular frequency $\omega$,
$E_s=\frac{1}{2}G\omega X^2$, where $G=2\pi L M_s/\gamma$ is the
gyrovector and $X$ the gyration radius of the core and $\Gamma_+= \eta
\alpha \omega$ \cite{guslienko06}. For the thick Py$_\text{B}$ layer
($L=15$~nm), $M_s=764$~emu.cm$^{-3}$,
$\gamma=1.87\cdot10^7$~rad.s$^{-1}$.Oe$^{-1}$ and the damping
coefficient $\alpha=0.008$ were determined by mechanical ferromagnetic
resonance \cite{naletov11}, and $\eta=1.7$ \cite{khvalkovskiy09}. From
micromagnetic simulations \cite{khvalkovskiy10a}, the radius of
gyration under bias conditions close to the experimental ones is found
to be $X\simeq 40$~nm (corresponding to $p=(X/R)^2\simeq0.1$). Hence,
at room temperature, $k_BT/E_s\simeq 0.003$ and
$\Gamma_+/(2\pi)=9.5$~MHz, which yields $\Delta F_g = 29$~kHz, a value
close to the narrow linewidth of 40~kHz observed at
$H_\text{ext}=-0.7$~kOe [Fig.\ref{fig:1}a], which thus almost
coincides with the intrinsic linewidth of our oscillator. In sum, the
absence of nonlinear broadening and the large energy stored in the
auto-oscillation in comparison to the thermal energy explain the low
minimal values ($<100$~kHz) of the linewidth in Fig.\ref{fig:2}a.

Before seeking further the origin of the variations of linewidth, we
first investigate possible physical causes of
quasi-isochronicity. Despite the fact that the nonlinearity $\nu$ of
our STNO is nearly zero, it exhibits a large frequency tunability,
$dF/dI_\text{dc} \simeq 7$~MHz/mA, see Fig.\ref{fig:2}b. This can be
explained by the linear contribution of the Oersted field to the
oscillation frequency. In fact, the tunability can be decomposed as
follows:
\begin{equation}
  \frac{dF}{dI_\text{dc}} = \frac{\partial F}{\partial p} \frac{\partial p}{\partial I_\text{dc}} + \frac{\partial F}{\partial I_\text{dc}} = N \frac{\partial p}{\partial I_\text{dc} } + A_\text{Oe} I_\text{dc} \, .
  \label{eq:tun}
\end{equation}
In the vortex state, the Oersted field participates to the confinement
potential of the vortex core \cite{khvalkovskiy09,dussaux12}, and for
our STNO parameters, it is predicted to be about $A_\text{Oe} \simeq
12.5$~MHz/mA. Therefore, a nonlinear frequency shift $N \neq 0$ is not
required to obtain a large tuning sensitivity in
Eq.(\ref{eq:tun}). Here, the measured tunability is comparable with
the one expected for the Oersted field alone. We attribute the
somewhat smaller experimental value to the fact that the
auto-oscillating mode does not only involve the thick Py layer, but
also the thin one \cite{locatelli11}, which is not taken into account
in the calculation of $A_\text{Oe}$.

For the gyrotropic mode, $N$ is the sum of the nonlinearities
$N_\text{ms}$ and $N_\text{Oe}$ of the magnetostatic and the Oersted
field confinement potentials, respectively. An analytical model
\cite{metlov13} predicts that the exact value of $N_\text{ms}$ is
positive and depends on the aspect ratio of the magnetic dot in the
vortex state, as confirmed by recent micromagnetic simulations
\cite{dussaux12} and experiments \cite{sukhostavets13}.  On the
contrary, $N_\text{Oe}$ is negative when the vortex chirality is
parallel to the Oersted field. Therefore, the two contributions can
compensate each other (see Fig.3 of Ref.\cite{dussaux12}): using our
STNO parameters, $N=N_\text{ms}+N_\text{Oe}=0$ for
$I_\text{dc}=23$~mA.  Experimentally, we find that $\nu \simeq 0$ is
robust from 18~mA up to 25~mA by analyzing the linewidths of harmonics
as in Fig.\ref{fig:2}a. By using Eq.(\ref{eq:harm}), we can be more
quantitative \cite{quinsat12}: in the full window of field and
current, we extract that $\nu<0.5$ and that $\Gamma_p$ ranges between
2 and 10~MHz. Hence, the intrinsically small $N$ with respect to
$\Gamma_p/p \simeq 10\cdot\Gamma_p$ in our STNO is probably
responsible for $\nu < 0.5$ \cite{braganca13}. Moreover, as mentioned
before, the auto-oscillating mode in our sample involves both the
thick and thin vortex Py layers with opposite core polarities. Several
non-conservative STT terms are thus involved in the dynamics
\cite{locatelli12}, which could be another cause of
quasi-isochronicity: micromagnetic simulations have indeed shown that
such terms can substantially reduce the nonlinear phase-amplitude
coupling \cite{gusakova11}.

\begin{figure}
  \includegraphics[width=\columnwidth]{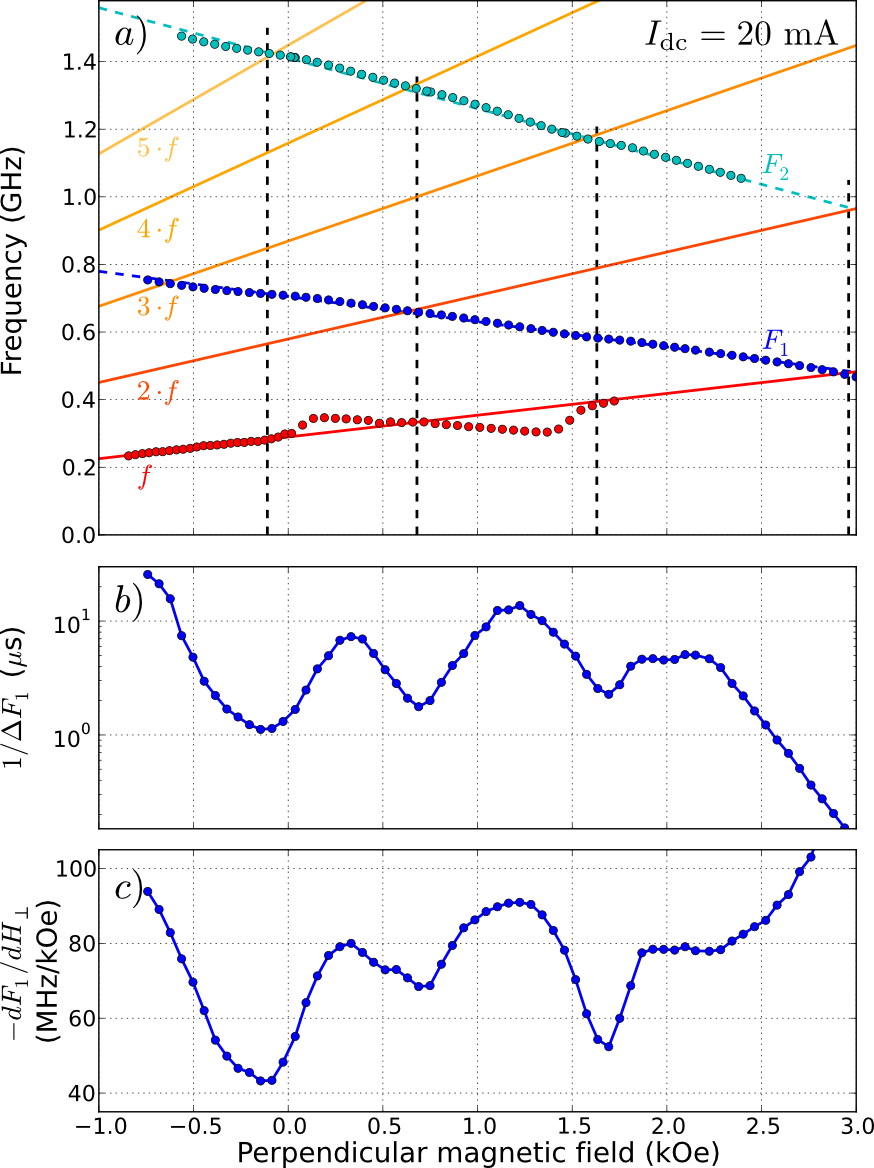}
  \caption{(Color online) (a) Blue dots: fundamental frequency $F_1$
    and harmonics $F_2$ of the auto-oscillating mode dominated by the
    thick layer (vortex core polarity $p_\text{B}=-1$) as a function
    of $H_\perp$. Red dots: frequency $f$ of the overdamped mode
    dominated by the thin layer (polarity $p_\text{A}=+1$). Red tone
    straight lines are guides to the eye and show successive harmonics
    of $f$. (b) Evolution of the inverse generation linewidth. (c)
    Slope of the oscillation frequency $F_1$ vs. $H_\perp$.}
  \label{fig:3}
\end{figure}

To elucidate the linewidth broadenings observed in Fig.\ref{fig:2}a,
it should be noticed that in addition to the gyrotropic mode dominated
by the thick Py$_\text{B}$ layer excited by STT, there is a gyrotropic
mode dominated by the thin Py$_\text{A}$ layer, which for
$I_\text{dc}>0$ is overdamped by STT \cite{locatelli12}. In order to
probe this mode, we use a microwave antenna deposited on top of the
sample which produces a microwave field $h_\text{rf}$ in the plane of
the Py layers \cite{naletov11,hamadeh12}, and detect the dc voltage
generated through the nanopillar when the vortex gyrotropic dynamics
is excited at resonance by $h_\text{rf}=3.6$~Oe \cite{kasai06}. We
have reported the measured resonant frequency associated to the thin
layer vortex mode as a function of $H_\perp$ in Fig.\ref{fig:3}a using
red dots. The detailed analysis of the observed behavior is out of the
scope of this Letter and will be reported separately. For the
following demonstration, it is sufficient to use a linear
approximation of it (red straight line). As expected, the gyrotropic
mode frequency $f$ dominated by Py$_\text{A}$ is lower and has an
opposite slope vs. $H_\perp$ than the one $F_1$ dominated by
Py$_\text{B}$, due to smaller thickness ($L_\text{A} < L_\text{B}$)
and opposite core polarity ($p_\text{A}=+1=-p_\text{B}$)
\cite{loubens09}. We also note that the linewidth of this mode ranges
between 80 and 100~MHz, which is about 8 times larger than its natural
linewidth, in agreement with the increase of relaxation of the thin
layer due to STT.

Using red tone lines in Fig.\ref{fig:3}a, we have plotted its
harmonics $n=2$ to 5 together with the measured fundamental frequency
$F_1$ and harmonics $F_2$ of the auto-oscillation signal (blue
dots). We have also reported in Figs.\ref{fig:3}b and \ref{fig:3}c the
inverse generation linewidth (logarithmic scale) and the slope
$-dF_1/dH_\perp$ of the oscillation frequency vs. $H_\perp$,
respectively. At magnetic fields where the frequencies of the
auto-oscillating and overdamped modes are commensurable, $pF_1=qf$,
with $p,q \in \mathbb{N}$ (dashed vertical lines in Fig.\ref{fig:3}a),
a decrease of $-dF_1/dH_\perp$ concomitant with a decrease of the
inverse linewidth is observed. We attribute this behavior to the
interaction between the eigenmodes of the STNO. When they cross each
other, their frequency dispersions soften \cite{pigeau12} and some
energy can be transferred from the auto-oscillating mode to the
overdamped mode. As a result, this additional channel of relaxation
leads to a decrease of the coherence time of the auto-oscillation
(this effect is not taken into account in Eq.(\ref{eq:deltaf}), which
was derived in a single mode approximation \cite{slavin09}). Here, the
dynamic dipolar interaction \cite{sugimoto11,keatley13} is an obvious
source of coupling between modes, but we stress that the dissipative
STT terms at play in the double vortex configuration can also be
important \cite{locatelli12}. Finally, since the coupling strength
depends on the difference in energy between the modes, we emphasize
that the STNO should better be operated under conditions where large
frequency gaps exist between the auto-oscillating mode and other modes
in the system.

In conclusion, we have presented a STNO based on coupled vortices
having its spectral purity and tuning sensitivity uncorrelated, which
is unusual for STNOs. We have demonstrated that it is
quasi-isochronous in a broad range of bias conditions, which
suppresses the most stringent cause of broadening and leads to high
spectral purity. The latter reaches its intrinsic value $Q>10000$ only
when the overdamped mode do not interact with the auto-oscillating
mode. We have also pointed out that $\nu \simeq 0$ can be due to the
compensation of the magnetostatic and Oersted field contributions to
the nonlinear frequency shift in our device. Interestingly for
applications, this does not prevent large tunability thanks to the
linear contribution of the current induced Oersted field to the
confinement potential of the vortex cores.


We are greatly indebted to A. N. Slavin for fruitful discussions and
for his support. This research was partly funded by the French ANR
(grant SPINNOVA ANR-11-NANO-0016) and the EU (FP7 grant MOSAIC
ICT-FP7-317950).


%

\end{document}